\documentclass[conference]{IEEEtran}
\IEEEoverridecommandlockouts
\usepackage{cite}
\usepackage{amsmath,amssymb,amsfonts}
\usepackage{algorithmic}
\usepackage{graphicx}
\usepackage{textcomp}
\usepackage{xcolor}
\usepackage{tabularx}
\usepackage{colortbl}
\usepackage{tikz}
\usepackage{array}
\usepackage{mdframed}
\usepackage{tikz}
\usepackage{booktabs}
\usepackage{multirow}
\usepackage{makecell}
\usepackage{stfloats}

\definecolor{roweven}{RGB}{240,240,240}
\definecolor{rowodd}{RGB}{255,255,255}

\usepackage{lipsum}  

\def\BibTeX{{\rm B\kern-.05em{\sc i\kern-.025em b}\kern-.08em
    T\kern-.1667em\lower.7ex\hbox{E}\kern-.125emX}}

\newenvironment{researchbox}{%
  \par\medskip\noindent%
  \begin{tikzpicture}%
    \node[inner sep=10pt, draw=black, fill=gray!10, line width=2pt, rounded corners=5pt]%
    \bgroup\minipage{\dimexpr0.46\textwidth-2\fboxsep-2\fboxrule\relax}%
}{%
    \endminipage\egroup;%
  \end{tikzpicture}%
  \par\medskip%
}

\begin{document}

\title{Leveraging Large Language Models for Predicting Cost and Duration in Software Engineering Projects}

\author{\IEEEauthorblockN{Justin Carpenter}
\IEEEauthorblockA{\textit{Department of Computer Science} \\
\textit{Boise State University}\\
Boise, Idaho, USA \\
Justincarpenter836@u.boisestate.edu}
\and
\IEEEauthorblockN{Chia-Ying Wu}
\IEEEauthorblockA{\textit{Department of Computer Science} \\
\textit{Boise State University}\\
Boise, USA \\
chiayingwu@u.boisestate.edu}
\and
\IEEEauthorblockN{Nasir U. Eisty}
\IEEEauthorblockA{\textit{Department of Computer Science} \\
\textit{Boise State University}\\
Boise, Idaho, USA \\
nasireisty@boisestate.edu}

}

\maketitle




\begin{abstract}
Accurate estimation of project costs and durations remains a pivotal challenge in software engineering, directly impacting budgeting and resource management. Traditional estimation techniques, although widely utilized, often fall short due to their complexity and the dynamic nature of software development projects. This study introduces an innovative approach using Large Language Models (LLMs) to enhance the accuracy and usability of project cost predictions. We explore the efficacy of LLMs against traditional methods and contemporary machine learning techniques, focusing on their potential to simplify the estimation process and provide higher accuracy. Our research is structured around critical inquiries into whether LLMs can outperform existing models, the ease of their integration into current practices, outperform traditional estimation, and why traditional methods still prevail in industry settings. By applying LLMs to a range of real-world datasets and comparing their performance to both state-of-the-art and conventional methods, this study aims to demonstrate that LLMs not only yield more accurate estimates but also offer a user-friendly alternative to complex predictive models, potentially transforming project management strategies within the software industry.
\end{abstract}


\section{Introduction}
Effective software cost estimation has always been a focal point for most companies in the field of software engineering. However, despite the existence of many traditional methods for cost estimation, the diversity and complexity of software development often render these traditional approaches inaccurate and time-consuming. For software development organizations, accurate cost estimation is paramount. It not only helps businesses avoid exceeding budgets but also facilitates effective resource allocation and utilization, ultimately enhancing product quality. However, many current methods of cost estimation suffer from various limitations, such as applicability constraints and low accuracy. Therefore, the core question of this study is: How can we leverage Large Language Models (LLMs) to improve the accuracy and reliability of software project cost estimation? Through addressing this question, we aim to provide software development organizations with more effective and practical cost estimation tools in the future, thereby enhancing companies' grasp and efficiency in the software development process. This question represents the primary focus of our research.

As time progresses, there is a growing awareness across various industries that effective software cost estimation is a crucial factor in reducing budget overruns in software projects. Proper software cost estimation can assist businesses in making reasonable investments in software and also enable software managers to gain a better understanding of the software being developed, thus facilitating effective supervision during the software development process. Software cost estimation not only aids in cost management but also contributes to enhancing product quality. Through the analysis of costs and historical project data, a better understanding of resource allocation and cost control can be achieved, consequently leading to improvements in software quality~\cite{marandi2015impact}. The better we understand software costs and quality, the better we can control them, and thus develop ideal products~\cite{boehm1988understanding}. More and more software development organizations are facing the intense challenges of global competition. The demand for high-quality products and new management paradigms has been proven to be crucial for success. In emerging markets, swift responses and accurate cost estimation are paramount for winning competitive bids and maintaining competitiveness~\cite{jorgensen2004review}.

There have been many different methods to estimate the cost and time a project will take to complete both before and during development. Traditional methods, while useful, tend to be time consuming and fall short due to the complex nature of software development. 
Typical means to estimate the cost occurs at 1\%-15\% of the project completion and the estimate accuracy varies from 30\% to 50\%.  We have seen previous approaches to utilize parametric models to create estimates where their accuracy varies from 50\% to 100\%, which is a significant improvement~\cite{hollmann_recommended_2009}. While parametric cost estimates present, more accurate results than capacity-factored models and analog models. They can tend to be difficult to work with and can struggle to branch out of the trained software area. 

Researchers have been employing LLMs in a wide range of software engineering tasks, data collection and pre-processing, optimization and evaluation techniques, and specific tasks. By utilizing LLMs in different applications, there have been some breakthroughs and some alternative approaches to solving previous problems, either better or worse than prior solutions~\cite{noauthor_large_nodate}. The versatility of LLMs allows for easy usage in this wide range of areas.

There is a wide need for automated project cost estimation that can be versatile and easy to use. We propose a novel solution which will utilize the versatile nature of LLMs to better predict project cost and duration more accurately. We theorize that while using LLMs over past automated approaches, we will be able to \textbf{1) Simplify the process of automated project cost estimating, 2) Utilize the versatility of LLMs to apply this approach to most software projects, and 3) Achieve higher accuracy in our predictions than prior approaches.} This will help improve the success rate and quality of software projects while reducing waste of time and cost. Our research will bring a new perspective to the academic community, exploring the potential application value of LLMs in software project management, thereby promoting research and development in related fields.

In recent years, the field of software engineering has undergone significant transformations with the widespread adoption of large language models. One key challenge in this evolution is the accurate prediction of project cost and duration. Inaccurate estimations can lead to substantial losses for businesses, highlighting the crucial importance of appropriate resource allocation in project management. This study focuses on exploring the application of large language models, investigating their potential in enhancing cost and efficiency predictions in project management within the realm of software engineering.

The exponential growth in publications related to LLMs in software engineering, indicating a significant interest in applying LLMs to software engineering tasks, such as analyzing software lifecycle data, code analysis, and providing just-in-time developer feedback. The potential of LLMs, aided by prompt engineering, to significantly improve software engineering efficiency by helping developers navigate and analyze vast amounts of project-related data, identify inconsistencies, and generate code with better security practices~\cite{zheng2023towards, noauthor_large_nodate}.

The more hands-on approach such as agile estimations have been more widely used in the course of software development, even with certain approaches which have been proven to be a more accurate approach. Roughly over 60\% of software developers have used agile in the past, and less than 10\% have used a prediction model approach~\cite{hidmi2017software}. By noticing, the estimation techniques chosen by most software developers do not solely rely on accuracy but on ease and understanding. This is why we propose the utilization of LLMs to allow developers to utilize an easy-to-use system in order to make an improved project cost estimation. 

In our study, we aim to explore the efficacy and usability of LLMs in the realm of software project management, particularly for estimating project costs and durations. Our research is structured around three fundamental questions, each aimed at uncovering different aspects of LLM application in this field.

\section{Research Questions}
\begin{enumerate}
    \item \textbf{Can LLMs outperform machine learning techniques in project cost and duration estimation?} This question seeks to ascertain whether LLMs, with their advanced capabilities in understanding and generating human-like text, can provide more accurate or reliable estimates compared to traditional machine learning methods.
    
    \item \textbf{Is an LLM approach to estimate the cost and duration of a project easier to utilize than the current approaches?} This inquiry focuses on the practical aspects of implementing LLMs in real-world settings, evaluating the ease of integration, user-friendliness, and accessibility of LLM-based models compared to conventional methods. Without the need for advanced features required to get an estimate like the machine learning models we observe.    
    \item \textbf{Why do companies still use traditional means to estimate projects cost?} Despite the advent of sophisticated predictive technologies, many organizations continue to rely on traditional estimation techniques. This question explores the reasons behind this preference, including factors such as the need for complicated features, resistance to change and the perceived reliability of traditional methods.
\end{enumerate}

Our proposed solution will be a simpler approach that will be compared against the current state-of-the-art, complicated and not popular, solutions and also compared against the traditional, easy and more popular solutions. By comparing against the traditional and the state-of-the-art we argue our solution will be more suited to real world uses with better outcomes than the traditional approach and easier to use than the current state-of-the-art approaches. In this work we will defend this claim by thoroughly testing our approach on 5 highly used real world datasets. 


\section{Related Works}
Predicting software development effort, size, and schedule are critical for planning and managing software projects effectively. Traditional software project estimation techniques, such as Constructive Cost Model (COCOMO)~\cite{boehm1987trw}, Delphi method~\cite{crisp1997delphi}, and agile estimations~\cite{cohn2005agile} have been a good starting point in many projects but have been known to be significantly off. The renowned software estimation technique, COCOMO, was originally developed by Barry Boehm in 1987. There have been numerous alternative models of COCOMO, such as COCOMO II~\cite{boehm1995cost}, which predict the effort and duration of a project based on inputs relating to the size of the resulting systems~\cite{kemerer_empirical_1987}.

Research studies such as those by Pospieszny et al.~\cite{pospieszny2018effective}, Aljahdali~\cite{aljahdali2010development}, and Khalid et al.~\cite{khalid2017approach} have also demonstrated the effectiveness of Machine Learning algorithms in software project effort and duration estimation. These studies have shown that models developed using algorithms like random forest, SVM, extreme reinforcement, logistic regression, decision tree, COA-Cuckoo and KNN can offer precise predictions for various software development projects. Furthermore, Abdulmajeed et al.'s study emphasizes in methods such as the K-Nearest Neighbours algorithm, Elman Neural Networks, and Cascade Neural Networks. These approaches have brought about high accuracy rates in predicting the costs required for developing software engineering projects.

A similar work done by Zaineb, Asma, and Nadia undertook a study evaluating software effort estimation utilizing machine learning regression methods unlike our approach with LLMs. They compared and tested four MLRM techniques (ABR, GBR, LinearSVR, RFR), and concluded that the Random Forest Regression (RFR) exhibited superior performance among them. Consequently, they recommended RFR for software enhancement effort estimation. Extracting insights from their research, we aim to utilize the gathered data to enhance a LLM classification~\cite{sakhrawi_software_nodate}.

In the research conducted by Bilge Başkeleş, Burak Turhan, Ayşe Bener, and others, the use of the COCOMO model in software engineering projects has demonstrated effectiveness on multiple fronts. Firstly, as a parametric model, the COCOMO model can estimate development time and effort based on various variables, contributing to improved accuracy in predicting project costs and timelines. Secondly, the study indicates that the COCOMO model exhibits high accuracy in similar projects, but it may have limitations for projects with different characteristics. This underscores the importance of utilizing large language models, as these models can adapt more flexibly to different types of software projects, enhancing prediction accuracy and flexibility. Through the examination of the COCOMO model, it can be demonstrated that adopting emerging technologies and approaches, such as large language models, contributes to the efficiency of cost and time predictions in software engineering projects~\cite{baskeles2007software}. 

Researchers V. Anandhi and R. Manicka Chezian have also conducted significant studies through the application of the COCOMO dataset, yielding important research findings~\cite{anandhi2014regression}. By analyzing software project-related data within the COCOMO dataset, including project size, complexity, and human resource allocation, they have constructed predictive models to assess the cost and time of software engineering projects. Leveraging historical data from the COCOMO dataset and utilizing large language models, the researchers conducted comprehensive analyses and predictions. They successfully established a predictive model capable of accurately forecasting the cost and time of software engineering projects.

Omar and Betul conducted a study on software effort estimation using machine learning methods. They developed a model utilizing Support Vector Machine (SVM) and K-Nearest Neighbor (k-NN) techniques, applied to the Desharnais and Maxwell datasets. Their combined approach yielded a 91\% accuracy rate for the Desharnais dataset and an 85\% accuracy rate for the Maxwell dataset~\cite{hidmi2017software} which is very high results given traditional results tend to be closer to 60\%. Our proposed solution will compare against traditional results, the more popular approach, and this state-of-the-art approach. We theorize that our approach will achieve higher results than traditional approaches and near the state-of-the-art approach which while being easier to implement for software developers, it will be a strong contribution in this field.

Researchers Fernando González-Ladrón-de-Guevara et al.~\cite{gonzalez2016usage} point out that the ISBSG dataset offers a rich resource for software engineering research, particularly in the domain of effort estimation models. They utilize the ISBSG dataset to investigate and develop models for estimating software project size, effort, duration, and cost. They emphasize that one of the key advantages of the ISBSG dataset lies in its extensive repository of data from completed software projects, providing insights into industry practices, methodologies, project outcomes, and tools. These rich and comprehensive insights enhance the quality of their research process, enabling them to improve the accuracy and reliability of their models through the utilization of this data. Therefore, we also anticipate utilizing the ISBSG dataset for our research to enhance the accuracy and reliability of the research process.

In a particular study, the advantages of utilizing the PROMISE dataset are highlighted. The strength of the PROMISE dataset lies in its provision of verifiable and reusable models for research in software engineering. Acting as an online repository, it facilitates the exchange of datasets among software engineering researchers and practitioners. Users have the option to either store their own datasets or access models shared by others, enabling them to compare results and draw support from prior studies~\cite{karim2017software}. Consequently, the use of the PROMISE dataset helps ensure the reproducibility and verifiability of research findings, thereby enhancing the credibility and value of the research.

While much effort and research has gone into the usage of machine learning to estimate the time and complexity of a given software project. Over the past decade, there has been major innovation and progress which has made predicting a projects cost possible. Even if these methods are not used more than a traditional approach, we will still use them to compare our approach due to them being the proposed state-of-the-art. 

In our work we will not only be focusing on traditional means to predict software cost and duration but also look at state of the art approaches that are not as commonly used due to the complexity nature of the approach. Different approaches to software project estimation, have utilized machine learning-based models which use ensemble methods and recursive feature selection for improved prediction accuracy. Some notable methods like software effort estimation (SEE) framework utilizing a support vector regression algorithm for feature extraction, a recursive elimination-based feature selection technique, and an enhanced random forest classifier for classification~\cite {hussain_enhanced_2021}.

In this research, we are going to use five publicly available datasets to fully test our approach against the prior state-of-the-art approaches and common agile approaches. The COCOMO, Desharnais~\cite{desharnais1989analyse}, The International Software Benchmarking Standards Group (ISBSG)~\cite{ISBSG2016}, QQDefects, and PROMISE datasets have been common datasets used by many prior software estimation survey sand will also be the primary for this work. These datasets are commonly used datasets in the field of software effort estimation and offer a rich foundation for software engineering research, encompassing a wide range of metrics crucial for effort estimation, defect prediction, and software quality analysis. These datasets include attributes like project size, development effort, team size, project duration, complexity measures, and defect counts, enabling comprehensive studies on productivity, project management effectiveness, and predictive modeling in software development. Their diverse metrics support various research inquiries into the factors influencing software project outcomes, serving as invaluable resources for validating theoretical models within the software engineering domain.

\section{Data}
There are different approaches to gather project cost datasets. Some which have been historically used but are considered out dated such as COCOMO and Desharnais which were made in the 1980s containing projects made in the past. We will use these in our testing resutlts but not to prove our model being working but to compair against our related works that have choosen to train and test with these models. 

\begin{table}
    \centering
    \caption{Project Data}
    \label{tab:datatable}
    \begin{tabularx}{0.3\textwidth}{|c|X|X|}
        \hline
        \rowcolor{gray!50}
        \textbf{Dataset}& \textbf{Projects} & \textbf{Features}  \\
        \hline
        \rowcolor{roweven}
        \textbf{Desharnais}                & 81           & 12        \\
        \hline
        \rowcolor{rowodd}
        \textbf{COCOMO}              & 93          & 23        \\
        \hline
        \rowcolor{roweven}
        \textbf{Maxwell}              & 62           & 26       \\
        \hline
        \rowcolor{rowodd}
        \textbf{ISBSG}              & 7518           & 264       \\
        \hline
    \end{tabularx}
\end{table}

\begin{figure*}[h!]
\centering
\includegraphics[scale=0.50]{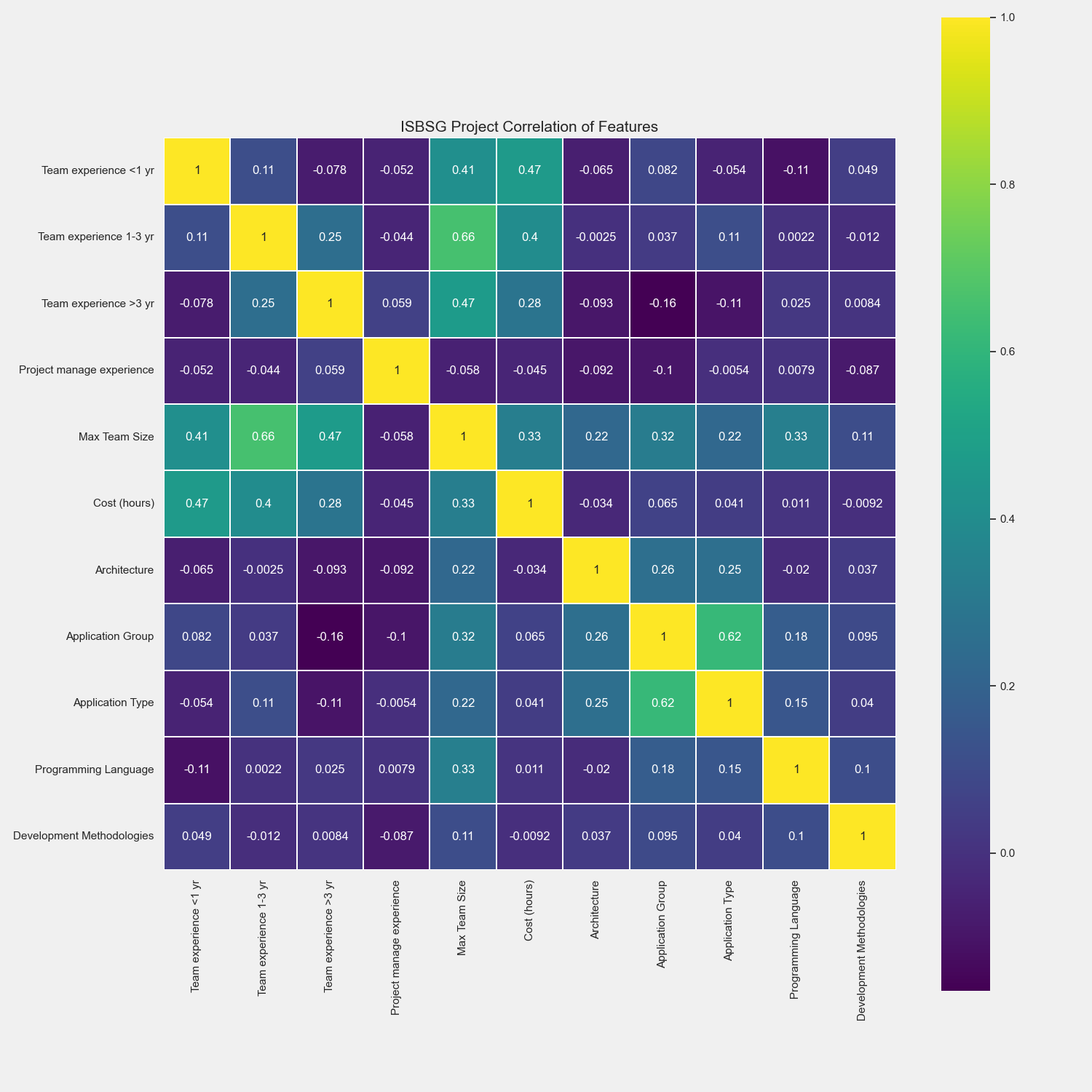}
\caption{Correlation of ISBSG Features with the normalized work effort (project cost)}
\label{fig:isbsg_corr}
\end{figure*}

Feature selection within our datasets were selected based on two reasons. The  first being correlated features, which we wanted features that had higher correlations with the estimated cost as these would prove most beneficial for our method. The second reason was more harmful to our results but to follow the reasoning of RQ1, we wanted to simplify the process of utilizing a LLM to estimate the cost of software projects. We settled on 10 features, shown in figure \ref{fig:isbsg_corr}. These features were selected as they are common among the beginning stages of a software development process. They also had high correlated features among the estimated cost. The other features were unique and likely would’ve benefited the LLM, they would’ve in turn complicated the process to utilize LLMs to predict the cost of the project. 

In our research utilizing the International Software Benchmarking Standards Group (ISBSG) dataset, we encountered the challenge of adapting structured software project data into a format suitable for fine-tuning the GPT-3.5 model. The ISBSG dataset, known for its comprehensive metrics on software development projects, required a transformation from a traditional tabular form into natural language prompts. To achieve this, we carefully selected ten key features from the dataset—such as project type, experience, architecture, team size, and technology used—that are particularly relevant for predicting project outcomes. Each entry in the dataset was then parsed to construct descriptive prompts that encapsulate the context of a software project in a concise statement.

This method of prompt engineering not only allows the LLM to understand and generate meaningful predictions based on the input features but also aligns with the model's capabilities to interpret and process natural language data. The conversion process involved meticulous formatting to ensure that the prompts are clear and structured in a way that maximizes the model's performance during the fine-tuning phase. This preparation was critical in leveraging the advanced AI's potential to generate accurate project cost and duration estimations, thereby enhancing the predictive accuracy of the model when applied to real-world software development scenarios.

In the process of creating prompts for fine-tuning the GPT-3.5 model using the ISBSG dataset, which comprises a comprehensive collection of 7518 software projects, we encountered significant challenges related to data completeness across the 10 features selected for our analysis. Initially, we utilized the entire dataset to generate prompts; however, it became evident that many entries contained numerous missing values (blanks), which could potentially compromise the quality of our model training.

\noindent \textbf{Prompt:} \\
\begin{verbatim}
What is the estimated cost of hours of a 
Project with the description: Architecture 
is Client server with 3.0 Number of 
developers with under 1 year of experiance, 
1.0 number of developers with 1 to 3 years 
of experiance, 0 Number of developers with 
over 3 years of experiance, Manager's years 
of experiance is 5.0, Business Application 
Application Group, Job, case, incident, 
project management; Application Type, 
Primary Programming Language is Java, Max 
Team Size of 4.0, Development Methodologies 
is Waterfall
\end{verbatim}

\noindent \textbf{Completion:} \\
\texttt{Estimated cost is: 1112.0 hours \\}

To address this issue and enhance the integrity of our training process, we stratified the dataset into subsets based on the number of missing values within these critical features. Our first refined subset included projects with at most five missing values among the selected features, resulting in a total of 3904 usable entries. This represented a substantial reduction but still retained a significant portion of the dataset, offering a balance between data completeness and sample size.

Further refining the data, we created another subset that included only those projects with at most three blanks, which dramatically reduced the pool to 228 entries. This subset aimed to prioritize data quality over quantity, ensuring more reliable input at the expense of a smaller dataset.

Finally, recognizing the need for the highest quality data for certain analyses, we compiled an exclusive subset of projects that had no missing data across the 10 selected features. This highly selective dataset comprised only 35 entries out of the original 7518, reflecting the stringent criteria applied. This smallest subset was critical for testing the model’s performance under the most ideal data conditions, even though it represented a minimal fraction of the entire dataset.

These tiered datasets allowed us to systematically explore the impact of data completeness on the model’s training efficacy and output accuracy, providing valuable insights into the trade-offs between data availability and quality in machine learning applications.
%
\section{Experiment}
%
%
%

The burgeoning complexity of software projects necessitates advanced methods for accurate cost estimation, particularly at the initial proposal stage where precise budgeting can significantly influence project approval and resource allocation. In response to this challenge, our research explores the potential of LLMs, specifically GPT-3.5, to revolutionize project cost estimation practices. The core of our experimental approach involves fine-tuning GPT-3.5 on a substantial portion of the ISBSG dataset, which provides a rich tapestry of project data ideally suited for this purpose.

Given the comprehensive nature of the ISBSG dataset, we allocated 80\% of the data for training and reserved 10\% for validation during the fine-tuning stage. This strategy ensures that the model not only learns from a diverse set of project scenarios but also adapts to unseen data, thereby enhancing its predictive robustness. The remaining 10\% of the ISBSG dataset serves as part of our testing framework, which also includes the Desharnais and COCOMO datasets. These datasets were selected to provide a broad base for testing, allowing us to assess the model's efficacy across different data structures and estimation paradigms.

This experimental setup is designed to rigorously test the hypothesis that LLMs can outperform traditional and current machine learning techniques in predicting software project costs at the proposal stage. By integrating LLMs into the early stages of project planning, we aim to provide project managers and stakeholders with a powerful tool that offers both high accuracy and ease of use, potentially setting a new standard in project cost estimation that can be easily adopted.

\subsection{Initial Attempts with Desharnais Datasets}
To begin our experimentation, we selected GPT-3.5 as our primary LLM due to its robust capabilities in handling diverse and complex data structures. Our initial approach involved utilizing the Desharnais dataset to fine-tune the model, adhering to an 80/20 training/testing split. However, the limited number of projects within the dataset proved insufficient for effective model training, resulting in performance that was subpar compared to a random baseline.

\subsection{Initial Attempts with Combined Datasets}
In an effort to enrich the training data, we combined the Desharnais dataset with the COCOMO dataset, maintaining the same 80/20 split for training and testing. This integration slightly improved the results, yet the outcomes were still not satisfactory, hovering around random choice levels in terms of accuracy and reliability.

\subsection{Transition to ISBSG Dataset}
Recognizing the need for a more substantial data foundation, we transitioned to the ISBSG dataset, which provided a broader array of project data suitable for comprehensive model training. We employed an 80/10/10 split (80\% for training, 10\% for testing, and 10\% for validation), specifically targeting this dataset to fine-tune the GPT-3.5 model. Subsequent tests were conducted using the Desharnais dataset, COCOMO dataset, and the 10\% validation split from ISBSG, ensuring a rigorous evaluation against comparable works.

\subsection{Detailed Data Handling for Fine-tuning}
The fine-tuning process was meticulously structured to accommodate the characteristics of the ISBSG dataset. Initially, we utilized the entire dataset to generate model prompts, integrating all 10 selected features. Given the prevalence of missing data, we created several sub-datasets to determine the impact of data completeness on model performance:
\begin{itemize}
    \item A subset containing 3904 entries, each with at most 5 missing values among the selected features.
    \item A more restricted subset of 228 entries with at most 3 missing values.
\end{itemize}
We did two fine tuning attempts which did use only the 3904 entries and the full dataset which when testing proved to have better results on the ISBSG dataset but worse on the desharnais and cocomo datasets. We strategically decided to utilize the full dataset with isolating the split with a 50\% split of the 228 entries amoung the training and testing splits. We believe this is due to the lack of features on the desharnais and cocomo datasets and in order to futher test RQ2 we wanted to show a LLM which can utilize a wide range of features. This is also the reason we opted to not use the smallest subset of 35 entries, which exhibited no missing values, due to its insufficient size for effective LLM training. We choose to use the full dataset but made sure to use a 50\% split of the 228 entries in the training and 50\% in the testing splits.

\subsection{Prompt Engineering and Model Training}
For each subset, we crafted specific prompts that encapsulated the project characteristics outlined in the ISBSG dataset, following the schema described earlier. This prompt engineering was crucial in adapting the tabular data into a format amenable to the LLM's training regime. Each subset underwent a separate fine-tuning process to tailor the model's understanding and predictive accuracy concerning the degree of data completeness.


\subsection{Performance Evaluation Measures}
To assess the accuracy of our LLM in estimating software project costs and durations, we utilized two primary statistical metrics: Mean Absolute Error (MAE) and Root Mean Square Error (RMSE). These measures provide insights into the average magnitude of the errors in predictions made by our model, without considering their direction (positive or negative).

\subsection{Mean Absolute Error (MAE)}

The Mean Absolute Error (MAE) is a measure of the difference between the actual and predicted values expressed as an average over the dataset. It is calculated as follows:

\begin{equation}
    \text{MAE} = \frac{1}{n} \sum_{i=1}^{n} |x_i - y_i|
\end{equation}

where $x_i$ represents the actual values, $y_i$ represents the predicted values, and $n$ is the number of observations. MAE provides a straightforward indication of average error magnitude and is particularly useful because it quantifies performance in the same units as the data, which aids in interpretation.

\subsection{Root Mean Square Error (RMSE)}

The Root Mean Square Error (RMSE) is a standard way to measure the error of a model in predicting quantitative data. Formally, it is defined as the square root of the average squared differences between the actual and predicted values:

\begin{equation}
    \text{RMSE} = \sqrt{\frac{1}{n} \sum_{i=1}^{n} (x_i - y_i)^2}
\end{equation}

Similar to MAE, $x_i$ and $y_i$ are the actual and predicted values, respectively, and $n$ is the number of observations. RMSE is sensitive to outliers and provides a higher weight to large errors, making it useful when large errors are particularly undesirable.

Both MAE and RMSE are critical in evaluating the prediction accuracy of our models, with RMSE being particularly useful when higher errors need to be penalized more severely than smaller ones. These metrics, together, offer a comprehensive view of the model performance across various aspects of the data's error distribution.

\section{Results}
\begin{table*}[bp]
    \centering
    \caption{Comparison of Root Mean Square Error (RMSE) results for baseline machine learning methods and the LLM across different datasets.}
    \begin{tabularx}{\textwidth}{|c|X|X|X|X|}
        \hline
        \rowcolor{gray!50}
        \textbf{Method}& \textbf{Desharnais}& \textbf{COCOMO} & \textbf{ISBSG}  \\
        \hline
        \rowcolor{roweven}
        \textbf{KNN Regression}        & \textbf{1755.11}  & 1127.50 & 15813.59   \\
        \hline
        \rowcolor{rowodd}
        \textbf{Linear Regression}     & 2680.12  & 1447.04 & 14309.82   \\
        \hline
        \rowcolor{roweven}
        \textbf{Support Vector Machine}& 2665.30  & 1475.71 & 14474.78   \\
        \hline
        \rowcolor{rowodd}
        \textbf{LLM}                   & 2522.21  & \textbf{537 .17} & \textbf{4848.76}    \\
        \hline
    \end{tabularx}
    \label{tab:datatable}
\end{table*}


\begin{figure}[h!]
\centering
\includegraphics[scale=0.17]{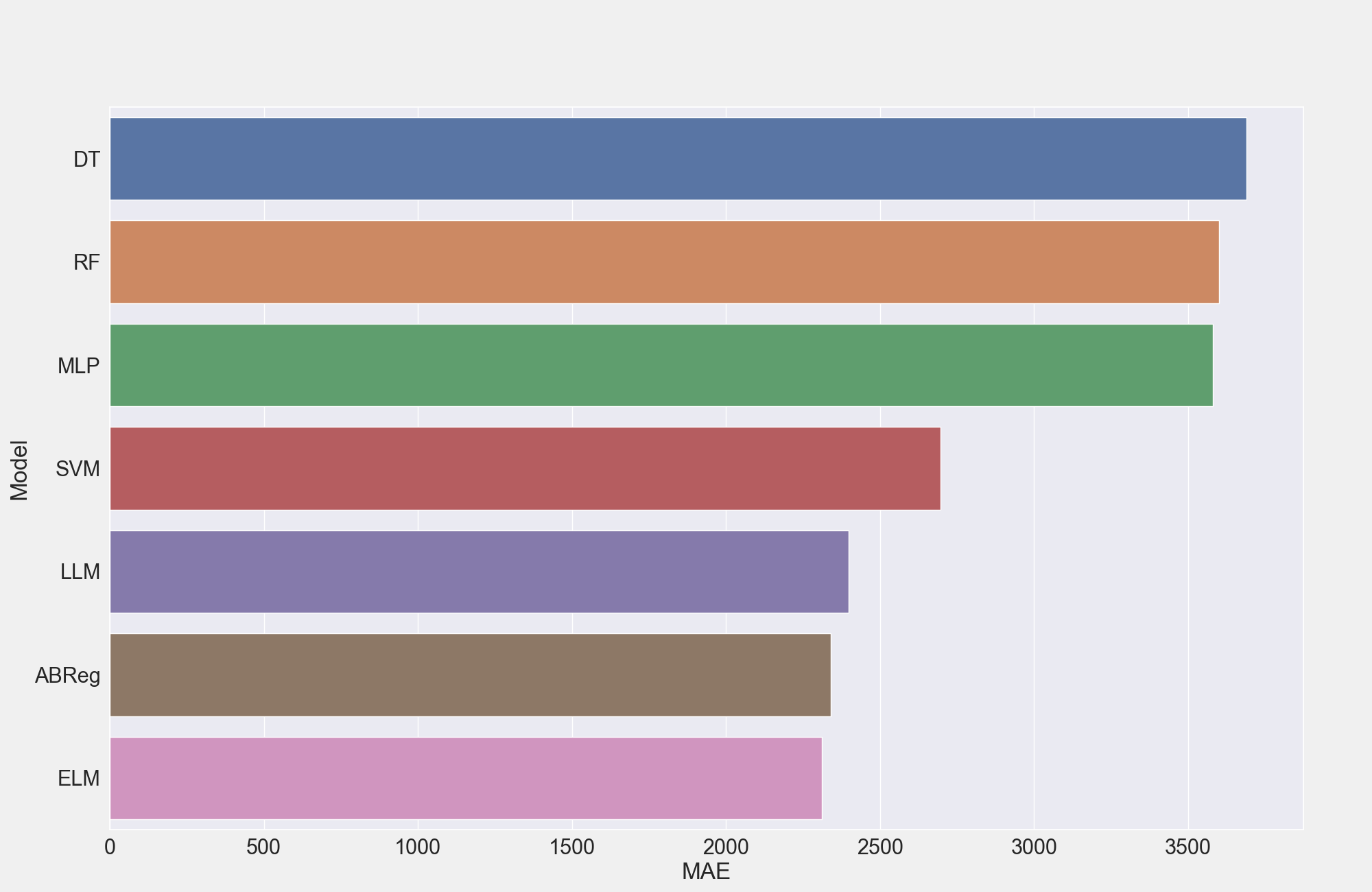}
\caption{Mean Absolute Error (MAE) on the Cost Estimation using Machine Learning and LLM's on the ISBSG dataset}
\label{fig:mae}
\end{figure}

In this section, our focus lies on evaluating the performance of LLMs in estimating software project costs and durations. We will commence by comparing the results of LLMs with traditional machine learning techniques and dive into the flexibility and accessibility of LLMs in project cost and duration estimation. Furthermore, we will discuss the research findings regarding LLMs' advantages in handling unseen data and their applications in various industries. Lastly, we will explore the reasons why companies still prefer using traditional methods for project cost estimation. Through these discussions, we aim to provide insights into the potential impact of LLMs in the field of project management and gain a deeper understanding of the challenges and limitations of existing methods.

\vspace{0.5em}{\textbf{\textit{RQ1: Can LLMs outperform machine learning techniques in project cost and duration estimation?}}}

\subsection{Comparison of LLMs with Machine Learning Techniques}

Our primary objective was to evaluate whether LLMs, specifically our fine-tuned GPT-3.5 model, could outperform traditional machine learning techniques in the domain of software project cost and duration estimation. The results, as detailed in Table~\ref{tab:datatable}, provide a comprehensive view of our findings across different datasets.

Initially, our comparison focused on traditional machine learning techniques, such as Support Vector Machines (SVM), K-Nearest Neighbors (KNN), and Linear Regression, which served as our baseline methods. The results indicate that while the LLM approach does not outperform the state-of-the-art machine learning techniques, it does exhibit superior performance compared to these baseline methods. This suggests a notable potential in LLMs for handling project estimation tasks, particularly because of their robustness in managing missing features—an area where traditional methods falter due to their dependency on complete feature sets.

In addition to the baseline methods we selected, we also choose five other machine learning approaches, namely Multi-Layer Perceptron (MLP)~\cite{murtagh1991multilayer}, Support Vector Machine (SVM)~\cite{drucker1996support}, Decision Tree (DT)~\cite{nie2011credit}, Random Forest (RF)~\cite{porru2016estimating},  Ada Boost Regression (ABReg)~\cite{solomatine2004adaboost}, and Extreme Learning Machine (ELM)~\cite{shukla2021extreme} to further extend our baseline on the ISBSG dataset. This includes supervised and unsupervised algorithms and a stat-of-the-art approach (ELM) in the effort to fully answer \textbf{RQ1}. In order to fully give the baseline machine learning approaches the best approach, we trained and tested them on the 3904 entries as they required no missing data. The LLM is the same training set we chose but the testing split is 10\% of the 3904 projects to closely compare against the other machine learning approaches.  As we can see in Table~\ref{fig:mae}, our approach is able to beat 4 out of the 5 machine learning approaches. The LLM is not better than the state-of-the-art method but it is notable to state our LLM approach has a 2398.2 MAE which is 3.7\% lower than the MAE of the state-of-the-art ELM which has a 2310.7 MAE. The ELM is trained on datasets which have no missing features and consequently must be used with no missing features. On the contrary the LLM has the potential, if trained using the same features and dataset, to more closely compare against the ELM.

\begin{researchbox}
  \textbf{Answer to RQ1:} LLM's can outperform some machine learning techniques but fall short against state of the art machine learning approaches.
\end{researchbox}

{\textbf{\textit{RQ2: Is an LLM approach to estimate the cost and duration of a project easier to utilize than the current approaches?}}}

\begin{figure*}[h!]
\centering
\includegraphics[scale=0.50]{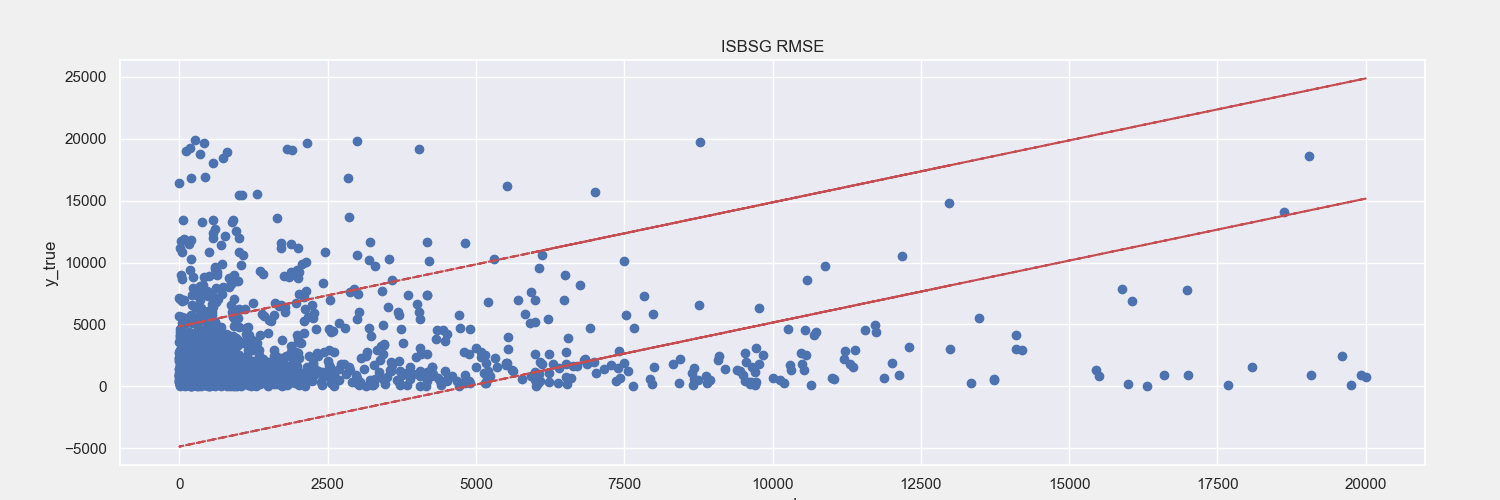}
\caption{True and Predicted cost with the RMSE score shown with the red line of ISBSG with outlier removed)}
\label{fig:isbsg_rmse}
\end{figure*}

\subsection{Flexibility and Accessibility of LLMs}

In evaluating the practicality of employing LLMs for estimating project costs and durations, we found LLMs to be distinctly more flexible than traditional machine learning approaches. A flaw with machine learning approaches tend to lie with requiring the features which the model was trained on, in order to get an estimation. This flexibility stems primarily from the LLM's ability to operate effectively even with incomplete data sets. Unlike conventional models that require a full set of features to function optimally, our LLM approach utilizes a set of 10 optional features. These features enhance the accuracy of predictions incrementally—the more features provided, the more precise the estimates. However, the model does not require any single feature to operate, making it highly adaptable to varied project scenarios.

\subsection{Deployment and Maintenance Advantages}

Moreover, the deployment and maintenance of LLMs offer significant advantages. Once tuned, LLMs can be easily distributed to end-users without the need for additional applications or the overhead associated with maintaining complex algorithms. This ease of distribution is particularly beneficial in environments where rapid deployment and ease of maintenance are critical. This has the potential to remove the reasoning why many developers tend to not adapt alternative approaches to estimate projects costs.

\subsection{Performance on Unseen Data}

To further test the robustness of our approach, we evaluated the fine-tuned LLM on datasets it had never been trained on, specifically the Desharnais and COCOMO datasets. Impressively, the LLM outperformed some machine learning methods that were explicitly trained on these datasets. This underscores the LLM's ability to generalize from limited data inputs to deliver reliable estimates, a feature particularly absent in traditional models that rely heavily on extensive feature sets.

\subsection{Comparative Advantage in Industry Applications}

Industry standards typically demand comprehensive information about projects before generating cost estimates. Which tend to be used after the project has begun. In contrast, our LLM approach can provide a preliminary estimate with minimal information and refine these estimates as more data becomes available. This scalability is a significant advantage over traditional methods, which do not offer the same flexibility to adapt to varying degrees of available information.

\begin{researchbox}
  \textbf{Answer to RQ2:} Yes. The findings indicate that the LLM approach not only simplifies the estimation process but also enhances the accessibility and adaptability of cost and duration estimation practices. This could herald a shift in how estimates are approached in the software industry, potentially leading to broader adoption of LLM-based techniques in project management. 
\end{researchbox}

{\textbf{\textit{RQ3: Why do companies still use traditional means to estimate projects cost?}}}

To comprehensively address this question, we will discuss it from two different perspectives and identify potential reasons. Firstly, we will discuss the advantages of traditional methods, highlighting their enduring value in certain contexts. Secondly, we will delve into the limitations of new methods, exploring the challenges they encounter despite their promise of higher accuracy. Through this analysis, we aim to provide a thorough understanding of why companies continue to rely on traditional methods for project cost estimation.

\textbf{a) Advantages of Traditional Methods:}  In certain contexts, traditional methods still hold distinct advantages.

\textit{1) Emphasis on Stakeholder Engagement and Communication:} Taking Agile estimation as an example, Agile methods have been widely adopted in software development, with over 60\% of software developers having utilized Agile methods. This indicates that traditional methods retain their unique value. Agile methodologies emphasize active customer engagement, effective communication, process simplification, fixed constraints, and user stories. These principles ensure a clear understanding of project requirements, reduce complexity, enhance efficiency, and facilitate cost estimation and development processes.

\textit{2) Reliability of Expert Judgment:} A study indicates that Expert Judgment, considered a traditional method, is widely used in the process of cost estimation. Cost estimators must make numerous assumptions and judgments about the costs of new products. While computerized cost models have reduced the need for Expert Judgment in many respects, they cannot completely replace it. This demonstrates the enduring applicability and importance of traditional methods even in the face of technological advancements~\cite{rush2001expert}. This technique relies on the extensive experience of cost estimation experts to estimate software costs. Naturally, experts' specialized knowledge in specific domains enables more precise cost estimations. Expert judgment proves particularly valuable in scenarios where effective and efficient data collection is restricted. In such cases, experts must make informed decisions based on various stringent criteria. The Delphi technique stands as an example of employing expert judgment~\cite{chirra2019survey}. Another study conducted an extensive review of expert estimation in software development effort and discussed 12 principles. The results indicated that expert estimates are more accurate in certain situations than formal models~\cite{jorgensen2004review}.

\textbf{b) Limitations of New Methods:} While new methods may promise higher accuracy, they encounter several challenges and limitations. Using a case study on technology transfer as an example, a software company experimented with a new software estimation method. Despite its higher accuracy, the organization did not fully embrace it. The reasons they gathered include perceptions of risk associated with new methods and the influence of organizational culture.

\textit{1) Risk and Uncertainty:}
One of the main reasons companies resist change is the perception of risks and uncertainties that new methods may bring. While new technologies may offer higher accuracy, the introduced uncertainty may make managers uneasy, especially during project execution. Using a case study on technology transfer as an example, a software company tried out a new software estimation method. Despite the new technology being more accurate, the organization did not fully adopt it. These are the reasons they have gathered: inadequate technical support, low user acceptance, difficulty in quantifying benefits, no single technique fits all problems, choose based on needs, research on development time estimation is limited~\cite{keung2004challenge}.

\textit{2) Influence of Organizational Culture:}
Organizational culture can also be a significant barrier. Over time, companies may have established a culture and processes based on traditional methods, making it challenging to enact change. Therefore, even though new methods may offer greater value, companies need to overcome internal cultural and organizational barriers to fully implement them. The impact of organizational culture can result in resistance towards new technologies, even when these technologies offer significant advantages in terms of efficiency and benefits~\cite{keung2004challenge}.

By further analyzing the advantages of traditional methods and discussing the challenges and limitations of new methods, we can gain a more comprehensive understanding of why companies still resort to traditional methods for project cost estimation.
\begin{researchbox}
  \textbf{Answer to RQ3:} Companies still rely on traditional methods for project cost estimation due to their emphasis on stakeholder engagement, communication, and the reliability of expert judgment. Meanwhile, the limitations of new methods, such as risks associated with change and the influence of organizational culture, hinder their full adoption.
\end{researchbox}

%
%



\section{Validity and Limitations}

\subsection{Financial and Resource Constraints}

One of the principal limitations encountered in this study relates to the constraints imposed by budgetary allocations. Financial restrictions significantly influenced our ability to extensively fine-tune the GPT-3.5 model. Optimal fine-tuning of large language models often requires substantial computational resources and financial investment, particularly when dealing with extensive parameter tuning and multiple dataset training cycles. Limited funding curtailed our ability to explore a broader array of model configurations and to potentially achieve more refined model performance.

\subsection{Dataset Limitations}

Additionally, our research was constrained by the version of the ISBSG dataset utilized. Due to budgetary limitations, we were restricted to an older version of the dataset. While this dataset provided a robust foundation for our initial analyses and model training, newer versions of the ISBSG dataset contain updated and possibly more varied project data that could enhance the model's learning potential and accuracy. Access to the latest dataset might have allowed for a more comprehensive understanding of current software project dynamics and potentially more accurate cost estimations.

\subsection{Impact on Study Outcomes}

These limitations are critical to acknowledge as they inherently affect the generalizability and applicability of our findings. The financial constraints and the use of an older dataset version may have prevented our models from achieving the performance levels that might be possible under more favorable circumstances. Future studies would benefit from securing additional funding to access updated datasets and more powerful computing resources, which would help in validating and possibly enhancing the conclusions drawn from our current research.

\section{Conclusion}

This research explored the potential of LLMs to revolutionize the estimation of software project costs and durations. Despite the advanced capabilities of state-of-the-art machine learning techniques, our findings reveal that LLMs hold significant promise due to their flexibility and ability to handle incomplete data sets effectively.

We systematically fine-tuned the GPT-3.5 model using a comprehensive set of data from the ISBSG dataset and tested its performance on multiple datasets, including the ISBSG testing set, Desharnais and COCOMO. While the LLM did not universally outperform all traditional machine learning models, it demonstrated superior performance over several baseline models and exhibited remarkable flexibility in dealing with sparse data inputs.

Our experiments confirmed that LLMs could provide accurate estimates even with minimal data, making them a practical tool for early-stage project planning. The ease of integration and user-friendliness of LLMs compared to conventional estimation techniques suggests a shift towards more adaptive and accessible tools in software project management.

\textbf{Future Work:} Looking ahead, further research should explore the integration of domain-specific knowledge into LLMs to enhance their predictive accuracy. Additionally, the development of hybrid models that combine the robust data handling capabilities of LLMs with the precision of algorithmic approaches could yield even more reliable estimation tools. Continuous refinement of the models and their training datasets will be crucial as we strive to meet the evolving demands of the software industry. Implementing alternative LLMs, with the growing amount of LLMs available, also have the potential to provide better results. 

In conclusion, while traditional methods remain entrenched in industry practices, the potential for LLMs to provide more efficient, adaptable, and user-friendly estimation tools is clear. Our study contributes to the growing body of knowledge demonstrating the viability of LLMs in practical applications, paving the way for their increased adoption in project cost estimation and beyond.


\bibliographystyle{IEEEtran}
\bibliography{main}

\end{document}